\documentclass[aip,amsmath,amssymb,jcp,nofootinbib,12pt]{revtex4-1}
\usepackage[utf8]{inputenc}
\usepackage{amsfonts}
\usepackage{amsmath}
\usepackage{amssymb}
\usepackage{amsthm}
\usepackage{appendix}
\usepackage{bbm}
\usepackage{chemformula}
\usepackage{enumerate}
\usepackage[T1]{fontenc}
\usepackage{datetime}
\usepackage{dcolumn} 
\usepackage{epigraph}
\usepackage[T1]{fontenc}
\usepackage{graphicx} 
\usepackage{hyperref}
\hypersetup{
    colorlinks=true,
    linkcolor=blue,
    filecolor=magenta,
    urlcolor=cyan,
}
\usepackage{cleveref}
\usepackage{soul} 
\usepackage{tabularx}
\usepackage{xcolor}
\usepackage{yfonts}

\newcommand{\erf}{\mathrm{erf}}
\newcommand{\erfc}{\mathrm{erfc}}
\newcommand{\bra}[1]{\ensuremath{\langle #1 \vert}}
\newcommand{\ket}[1]{\ensuremath{\vert #1  \rangle}}

\newcommand{\R}{\mathbb{R}}  

\makeatletter
\newcommand{\thickbar}{\mathpalette\@thickbar}
\newcommand{\@thickbar}[2]{{#1\mkern1.5mu\vbox{
  \sbox\z@{$#1\mkern-1.5mu#2\mkern-1.5mu$}%
  \sbox\tw@{$#1\overline{#2}$}%
  \dimen@=\dimexpr\ht\tw@-\ht\z@-.8\p@\relax
  \hrule\@height.8\p@ 
  \vskip\dimen@
  \box\z@}\mkern1.5mu}
}
\makeatother
\renewcommand*{\bar}{\thickbar}

\begin{document}
\title
{Correcting models with long-range electron interaction using
generalized cusp conditions}

\author{Andreas Savin$^\ast$}
\affiliation{Laboratoire de Chimie Théorique, CNRS and Sorbonne University
4 place Jussieu, 75252 Paris cedex 05, France}
\email{andreas.savin@lct.jussieu.fr}
\author{Jacek Karwowski}
\affiliation{Institute of Physics, Faculty of Physics, Astronomy and 
Informatics, Nicolaus Copernicus University, Grudziadzka 5, 87-100 Toru\'n, 
Poland}
\email{jka@umk.pl}


\begin{abstract}
\label{abs}
Sources of energy errors resulting from the replacement of the physical
Coulomb interaction by its long-range $\erfc(\mu\,r)/r$ approximation are
explored. It is demonstrated that the results can be dramatically improved
and the range of $\mu$ giving energies within chemical accuracy limits
significantly extended, if the generalized cusp conditions are used to
represent the wave function at small $r$.  The numerical results for
two-electron harmonium are presented and discussed.\\
\end{abstract}

\maketitle

\section{The problem to be solved}
\label{s-prsol}
We have a model system, $H(\mathbf{R};\mu)$, and a 
corresponding Schr\"odinger equation,
\begin{equation}
\label{01}
H(\mathbf{R};\mu)\Psi(\mathbf{R},\pmb{\sigma};\mu)= 
E(\mu)\Psi(\mathbf{R},\pmb{\sigma};\mu).
\end{equation}
The system is composed of $N$ electrons confined by an external potential,
$\mathbf{R}$ and $\pmb{\sigma}$ stand, respectively, for their
orbital and spin coordinates.  All quantities characterizing the system
(e.g. energy or wave function) depend on the external potential, but we
show this dependence explicitly only when the form of this potential is
specified (e.g.  the dependence on $\omega$ in the section~"The model system".

The interaction between electrons is described by a $\mu$-dependent model
potential $v_\mathrm{int}(r;\mu)$:
\begin{itemize}
\item $\mu=0$:  there is no interaction between electrons, so
$v_\mathrm{int}(r;0)=0$,
\item $\mu=\infty$: we have the physical, Coulomb interaction, so 
$v_\mathrm{int}(r;\infty)=1/r$,
\item $\mu\in(0,\infty)$: we choose  
\begin{equation}
\label{02} 
v_\mathrm{int}(r;\mu)=w(r;\mu)=\frac{\erf(\mu r)}{r},
\end{equation}
where $r=r_{12}=\left|\mathbf{r}_1-\mathbf{r}_2\right|$. 
Exploring other forms of interaction may be both interesting
and useful as, for example, in ref~\citenum{GonAyeKarSav-16}.
\end{itemize}
To simplify the notation, we drop $\mu$ when $\mu=\infty$. 

We assume that the solutions of eq~\eqref{01} are accessible for selected finite
values of $\mu$.  However, we are not interested in the model system energy,
$E(\mu)$.  We aim at determining $E$ corresponding to the physical
interaction.  Stated differently, we are interested in
\begin{equation}
    \label{03}
    \thickbar{E}(\mu)=E-E(\mu)\,\equiv\,-\Delta_\mathrm{0}E(\mu),
\end{equation}
where $\Delta_\mathrm{0}E(\mu)$ is referred to as the {\em error of the energy of
the model system}.

\begin{figure}
    \centering
    \includegraphics{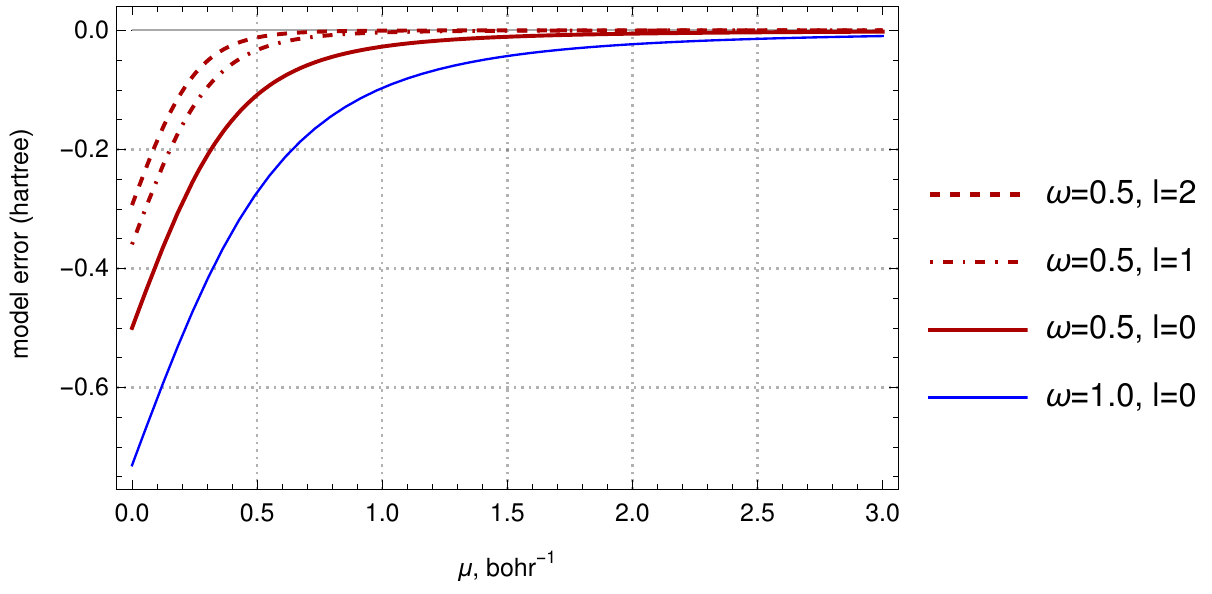}
    \caption{Error of the energy of the model system,
    $\Delta_\mathrm{0}E(\mu)$, for the lowest energy state of harmonium with
    $\omega=1/2$ (red curves), and $\omega=1$ (thin blue curve), for
    $\ell=0$, full curves; $\ell=1$, dot-dashed curve; $\ell=2$, dashed
    curve.}
    \label{fig:known-e-bar}
\end{figure}
To get an idea about the change of $\thickbar{E}(\mu)$ with respect to $\mu$,
we show in Figure~\ref{fig:known-e-bar} some situations where $E$ is known
for arbitrary precision. By construction, $\thickbar{E}(\infty)=0$.  As $\mu$
decreases, the interaction weakens and disappears for $\mu=0$.  This effect,
not compensated by any change in the external potential, leads to the
absolute values of $\thickbar{E}(\mu)$ increasing with decreasing $\mu$ and
becoming very large for sufficiently small $\mu$.

In this paper we explore how much one can lower the values of $\mu$ and, by
correcting the model, still retain approximations of $\thickbar{E}$ within
the {\em chemical accuracy} ($\pm\,1\,\mathrm{kcal/mol}$) error bars.

\section{Correcting models}
\label{s-cmod}
\subsection{Energy extrapolation -- a historic solution}
\label{ss-enex}
A way proposed in ref~\citenum{Sav-JCP-11} is to expand $\thickbar{E}(\mu)$ 
in the following basis,
\begin{equation}
\label{04}
    \thickbar{E}(\mu) \approx \sum_{k=1}^M \bar{e}_k \chi_k(\mu).
\end{equation}
Here $\chi_k(\mu), \, k=1,\dots, M$ are some basis functions, and
$\bar{e}_k$ are coefficients to be determined.  There are different ways to
determine these coefficients, once the basis functions are given.  One
option is to obtain them from the derivatives of $E(\mu)$ with respect to
$\mu$.  This corresponds to (generalized) Taylor expansions, or to
perturbation theory.  Another possibility is calculating $E(\mu)$ for
several values of $\mu$, and then matching the results to the expansion. 
This method is called {\em energy extrapolation}~\cite{Sav-JCP-11}: from
model information, we aim to reach the physical result.

One way to achieve our aim is to introduce more parameters into the
Hamiltonian~\cite{Sav-JCP-20}, e.g.,
\begin{equation}
\label{05}
H(\mathbf{R};\lambda,\mu)=H(\mathbf{R};\mu)+\lambda
\left[H(\mathbf{R})-H(\mathbf{R};\mu)\right].
\end{equation}
The eigenvalue and the corresponding eigenfunction of 
$H(\mathbf{R};\lambda,\mu)$ are, respectively, $E(\lambda,\mu)$ and
$\Psi(\mathbf{R},\pmb{\sigma};\lambda,\mu)$ --
notice that $E(\mu)=\left.E(\lambda,\mu)\right|_{\lambda=0}$.  
In the Hamiltonian~\eqref{05} the interaction potential is
\[
v_\mathrm{int}(r;\lambda,\mu)=w(r;\mu)+\lambda\,\bar{w}(r;\mu),
\]
where
\begin{equation}
\label{06}
\bar{w}(r;\mu)=\frac{1}{r}-w(r;\mu)=\frac{1-\erf(\mu r)}{r}= 
\frac{\erfc(\mu r)}{r}.
\end{equation}
Therefore,
\begin{equation}
\label{07}
v_\mathrm{int}(r;\lambda,\mu)=(1-\lambda)\frac{\erf(\mu r)}{r}+
\frac{\lambda}{r}
\end{equation}

We see that we can reach the physical result either with $\lambda=1$, or
with $\mu=\infty$.  "Shooting" from different points to the same target may
simplify our task.  However, this is not further discussed in this paper.

Energy extrapolation has an important problem: we do not know how to choose
$\chi_k(\mu)$.  What makes the problem worse is that we are not willing to
use many basis functions.  Ideally, we should use a single function, that is
to perform a single model calculation, $M=1$ in eq~\eqref{04}.

\subsection{The adiabatic connection}
\label{s-adconn} 

For $\left\|\Psi(\mathbf{R},\pmb{\sigma};\lambda,\mu)\right\|=1$, 
the Hellmann-Feynman theorem yields  
\begin{equation}
\label{08}
\partial_\lambda E(\lambda,\mu)=
\bra{\Psi(\mathbf{R},\pmb{\sigma};\lambda,\mu)}\partial_\lambda 
H(\mathbf{R};\lambda,\mu) 
\ket{\Psi(\mathbf{R},\pmb{\sigma};\lambda,\mu)}
=\langle\bar{w}(\lambda,\mu)\rangle,
\end{equation}
where
\begin{equation}
\label{09}
\langle\bar{w}(\lambda,\mu)\rangle=
\bra{\Psi(\mathbf{R},\pmb{\sigma};\lambda,\mu)}\bar{W}(\mathbf{R};\mu)
\ket{\Psi(\mathbf{R},\pmb{\sigma};\lambda,\mu)},
\end{equation}
and
\begin{equation}
\label{10}
\bar{W}(\mathbf{R};\mu)\equiv
H(\mathbf{R})-H(\mathbf{R};\mu)=\sum_{1\le i < j \le N} 
\bar{w}(r_{ij};\mu),
\end{equation}

By integrating eq~\eqref{08} over $\lambda$ one obtains that
\footnote{Notice that $E$ could have also been obtained by
integration over $\mu$, as $E=E(1,\mu)=E(\lambda,\infty)$.}
\begin{equation}
\label{11}
\thickbar{E}(\mu)=\int_0^1 \langle\bar{w}(\lambda,\mu)\rangle\,d\lambda. 
\end{equation}
The integrand in eq~\eqref{11}, an integral over $3N$-dimensional 
configuration space and $2^N$-dimensional spin space, can be reduced to
a one-dimensional radial integral. Exploiting the antisymmetry
of the wave function and integrating over spin, and over 
coordinates of electrons $3,4,\ldots,N$, yields \cite{lowdin-1955} 
\begin{equation}
\label{12}
\langle\bar{w}(\lambda,\mu)\rangle=\int_{\R^6} \bar{w}(r_{12};\mu)
\Gamma_{\lambda,\mu}(\mathbf{r}_1,\mathbf{r}_2)d\mathbf{r}_1d\mathbf{r}_2,
\end{equation}
where
\[
\Gamma_{\lambda,\mu}(\mathbf{r}_1,\mathbf{r}_2)=\binom{N}{2} 
\sum_{\sigma_1, ..., \sigma_N} \int_{\R^{3N-6}}
\left|\Psi(\mathbf{R},\pmb{\sigma};\lambda,\mu)\right|^2
d\mathbf{r}_3 d\mathbf{r}_4\cdots d\mathbf{r}_N
\]
is the diagonal part of the second-order reduced density matrix, 2RDM, 
corresponding to $\Psi(\mathbf{R},\pmb{\sigma};\lambda,\mu)$; the sum
is extended over spin coordinates of all electrons. 

After introducing the relative-motion variables
\begin{equation}
\label{13}
\mathbf{r}=\mathbf{r}_1-\mathbf{r_2},\;\;\;\;
\mathbf{r}^+=\frac{\mathbf{r}_1+\mathbf{r_2}}{2}, 
\end{equation}
performing integration over $\mathbf{r}^+$, and expressing $\mathbf{r}$ in
spherical coordinates, $\mathbf{r}(r,\theta,\phi)$, eq~\eqref{12} can be
rewritten as
\begin{equation}
\label{14}
\langle\bar{w}(\lambda,\mu)
\rangle=\int_{\R^6}\bar{w}(r;\mu)
\Gamma_{\lambda,\mu}(\mathbf{r},\mathbf{r}^+)d\mathbf{r}d\mathbf{r}^+=
\int_{\R^3}\bar{w}(r;\mu)\gamma(\mathbf{r} ;\lambda,\mu)d\mathbf{r},
\end{equation}
where $d\mathbf{r}=r^2\,dr\,\sin\theta\,d\theta\,d\phi$,
$r=r_{12}=\left|\mathbf{r}_1-\mathbf{r}_2\right|$, and
$\gamma(\mathbf{r;\lambda,\mu})$ is the diagonal part of the first-order
reduced density matrix, 1RDM.  Since in the coordinate space $\bar{w}(r;\mu)$
depends on the radial coordinate $r$ only, we can do the spherical
averaging.  In effect the integrand of eq~\eqref{11} is simplified to
\begin{equation}
\label{15}
\langle\bar{w}(\lambda,\mu)\rangle=\int_0^\infty\,\bar{w}(r;\mu)
\tilde{\gamma}(r;\lambda,\mu)r^2dr,  
\end{equation}
where 
\[
\tilde{\gamma}(r;\lambda,\mu)=\int_0^{2\pi}\int_0^\pi
\gamma(\mathbf{r};\lambda,\mu)\sin\theta\,d\theta\,d\phi.
\]
The adiabatic connection defined in eq~\eqref{11} carries no practical
information, as it requires the knowledge of
$\langle\bar{w}(\lambda,\mu)\rangle$ for all values of $\lambda$ while in
the present approach it is known only for $\lambda=0$.

At the limit of $r\rightarrow 0$, $\bar{w}(r;\mu)\sim 1/r$.  Therefore, also
for large $\mu$, $\bar{w}(r;\mu)$ is non-negligible if $r$ is small enough. 
As is shown hereafter, the information necessary for correcting the model at
small $r$ can be derived from the generalized cusp conditions (GCC).

\subsection{Generalized cusp conditions}
\label{ss-gcc}

The information about the behavior of the wave function in the vicinity of
the coalescence point, i.e., for $r=r_{12}\ll{1}$, can be derived from
general properties of the Schr\"odinger equation at $r\rightarrow 0$.  In
general, the approach is based on the expansion of the wave function and of
the potential as the power series of $r$ and deriving conditions that have to
be fulfilled by the expansion coefficients in order to retain the
consistency of the Hamiltonian eigenvalue problem.  The simplest and most
commonly known is the Kato's cusp condition~\cite{kato}, which can be
derived from the requirement that in the case of the electrostatic
interaction the local energy at $r=0$ is nonsingular.  Higher order
(generalized) cusp conditions can be obtained from the demand that the local
energies generated by powers of the Hamiltonian are nonsingular 
(energy-independent conditions) and ratios of the local energies generated 
by the consecutive powers are constant (energy-dependent
conditions)~\cite{jk+as}.  Alternatively one can use the expansion of the
wave function in powers of $r$ and require that the Schr\"odinger equation
is statisfied~\cite{kurokawa1,kurokawa2,kurokawa3}. Both approaches
are equivalent, but the conditions derived from the former one, though
more complicated, have more transparent physical meaning.
 
In this paper the GCCs are applied to describe the $r$-dependence of the
wave function in the area of small $r$, where the model interaction
potential departs from the physical one.

\section{The model system}
\label{s-modsys}

The simplest nontrivial model system containing one pair of
electrons is composed of three particles: two electrons interacting by a
repulsive model potential, and a third particle, "nucleus", which interacts
with electrons by an attractive force.  Commonly known examples of such
systems are helium atom - the nucleus attracts electrons by the Coulomb
force, and harmonium (Hooke atom) - the nucleus attracts electrons by the
Hooke force.  After the separation of the center of mass, the system is
reduced to two interacting particles in an external potential
$v_\mathrm{ext}$. In the case of harmonium,
\begin{equation}
\label{16}
v_\mathrm{ext}(r_1,r_2;\omega)=
\frac{\omega^2}{2}\left(r_1^2+r_2^2\right).
\end{equation}
The potential depends on a parameter, $\omega$, which defines the strength 
of the confinement. In the case of quantities that depend on this
potential, the dependence on $\omega$ is explicitly shown. For example,
$E(\omega,\mu)$ stands for the special case of $E(\mu)$, corresponding
to the external potential~\eqref{16}.    

To our knowledge, harmonium is the only bound system containing a
pair of interacting electrons for which the Schr\"odinger equation is known
to be separable.  Apart from the three-dimensional free-particle equation
describing the motion of the center of mass, the two-particle Schr\"odinger
equation for harmonium is separable into six one-dimensional equations - five
are exactly solvable, and the sixth one can be solved numerically to an
arbitrary precision (for some specific values of $\omega$ it is also
solvable analytically).  It is important to note that the separability holds
for all forms of the interaction potential including the form given by
$v_\mathrm{int}(r;\lambda,\mu)$.~\footnote{In principle the term 'harmonium'
refers to two confined Coulomb-interacting electrons.  In this paper we
extend this term to the case of model potentials $v_\mathrm{int}$.}
Therefore, harmonium is particularly suitable for pilot studies of the
consequences of using various non-Coulombic forms of the interaction
potentials.  Motivated by these observations, at this stage, we explore our
problem using harmonium as the model system.

The Schr\"odinger equation for harmonium
\begin{equation}
\label{17}
\left[T(\mathbf{r}_1,\mathbf{r}_2)
+v_\mathrm{ext}(r_1,r_2;\omega)+v_\mathrm{int}(r;\lambda,\mu)
-\mathcal{E}(\mathfrak{p})\right]
\Psi(\mathbf{r}_1,\mathbf{r}_2;\mathfrak{p})=0,
\end{equation}
where $T(\mathbf{r}_1,\mathbf{r}_2)$ is two-particle kinetic energy
operator, depends on three parameters, collectively denoted
$\mathfrak{p}=\{\omega,\lambda,\mu\}$, and
$\Psi(\mathbf{r}_1,\mathbf{r}_2;\mathfrak{p})$ is the orbital part of the 
two-electron wave function.  After transformation~\eqref{13},
eq~\eqref{17} can by split into two spherically-symmetric equations.  The
first one depends on the interaction potential and describes the relative
motion of electrons:
\begin{equation}
\label{18}
\left[-\Delta_\mathbf{r}+v(r;\mathfrak{p})-
E(\mathfrak{p})\right]
\psi_\mathrm{rel}(\mathbf{r};\mathfrak{p})=0,
\end{equation}
where
\begin{equation}
\label{19}
v(r;\mathfrak{p})=
\frac{\omega^2\,r^2}{4}+v_\mathrm{int}(r;\lambda,\mu)
=\frac{\omega^2\,r^2}{4} +
(1-\lambda)\frac{\erf(\mu r)}{r}+\frac{\lambda}{r}.
\end{equation}
The second equation describes the motion of the center of mass
of the electron pair in the external potential:
\begin{equation}
\label{20}
\left[-\frac{\Delta_{\mathbf{r^+}}}{4}+\omega^2(r^+)^2
-\mathfrak{E}(\omega)\right]\psi_\mathrm{cm}(\mathbf{r}^+;\omega)=0,
\end{equation}
where $r^+=|\mathbf{r^+}|$.
By construction, we have
\begin{equation}
\label{21}
\mathcal{E}(\mathfrak{p})=E(\mathfrak{p})
+\mathfrak{E}(\omega),\;\;\;\;\;
\Psi(\mathbf{r}_1,\mathbf{r}_2;\mathfrak{p})=
\psi_\mathrm{rel}(\mathbf{r};\mathfrak{p})\,
\psi_\mathrm{cm}(\mathbf{r}^+;\omega).
\end{equation}
The interaction potential appears only in eq~\eqref{18}. So, for our
study, we deal with this equation only. The potential is spherically 
symmetric. Therefore,
\begin{equation}
\label{22}
\psi_\mathrm{rel}(\mathbf{r};\mathfrak{p})=\psi_\ell(r;\mathfrak{p})
Y_{\ell m}(\theta,\phi),
\end{equation}
where $\psi_\ell(r;\mathfrak{p})$ is determined by the radial equation 
\begin{equation}
\label{23}
\left[-\frac{d^2}{dr^2}+\frac{\ell(\ell+1)}{r^2}+
v(r;\mathfrak{p})-E(\mathfrak{p})\right]
\left[r\,\psi_\ell(r;\mathfrak{p})\right]=0.
\end{equation}

The two-electron wave function,
$\Psi(\mathbf{r}_1,\mathbf{r}_2;\mathfrak{p})$, symmetric/antisymmetric with
respect to the transposition of $(\mathbf{r}_1,\mathbf{r}_2)$ correspond to
singlet/triplet.  As one can see, singlet states correspond to the even
parity (even $\ell$) spherical harmonics in eq~\eqref{22}, and triplet
states - to the odd ones.

\subsection{Generalized cusp conditions for the model system}
\label{ss-gccmod}
Using~\cite{ryzhik}
\[
\frac{\erf(\mu r)}{r}=\frac{2\mu}{\sqrt{\pi}}
\left(1-\frac{(\mu r)^2}{3\cdot 1!}+\frac{(\mu r)^4}{5\cdot 2!}-
\frac{(\mu r)^6}{7\cdot 3!}+\cdots\right)    
\]
one can expand the potential~\eqref{19} as
\begin{equation}
\label{24}
v(r;\mathfrak{p})=\sum_{i=-1}^\infty v_i(\mathfrak{p})\,r^i,
\end{equation}
with
\begin{equation}
\label{25}
v_{-1}=\lambda,\;\;v_{0}=(1-\lambda)\frac{2\mu}{\sqrt{\pi}},\;\;
v_1=0,\;\;v_2=\frac{\omega^2}{4}-(1-\lambda)\frac{2\mu^3}{3\sqrt{\pi}},
\;\;v_3=0,\ldots.
\end{equation}

The wave function, for small $r$, can be represented by the following 
power series 
\begin{equation}
\label{26}
\psi_\ell(r;\mathfrak{p})\approx r^\ell\sum_{k=0}^K c_k(\mathfrak{p})
r^k = c_0(\mathfrak{p})\,r^\ell \sum_{k=0}^K \widetilde{c}_k(\mathfrak{p})\,r^k,
\end{equation}
where $\widetilde{c}_k(\mathfrak{p})=c_k(\mathfrak{p})/c_0(\mathfrak{p})$.
General formulas for GCC are given in
refs~\citenum{kurokawa1,kurokawa2,kurokawa3,jk+as}. Here we give equations
defining $\widetilde{c}_k$ for $k\le 4$:
\begin{eqnarray}
\label{eq:GCC}
\nonumber
A_1\widetilde{c_1}+v_{-1}=0,,\\
\nonumber
A_2\widetilde{c_2}+v_{-1}\widetilde{c_1}-\epsilon=0,\\
\label{27}
A_1A_3\widetilde{c_3}+(A_1+A_2)v_{-1}\widetilde{c_2}
+v_{-1}^2\widetilde{c_1}+A_1v_1=0,\\
\nonumber
A_2A_4\widetilde{c_4}+(A_2+A_3)v_{-1}\widetilde{c_3}
+v_{-1}^2\widetilde{c_2}+A_2v_1\widetilde{c_1}+(v_{-1}v_1+A_2v_2
-\epsilon^2)=0,\\
\nonumber
\cdots\; \cdots\; \cdots
\end{eqnarray}
where $A_i=-i(2\ell+i+1)$, and $\epsilon=E-v_0$.  Coefficients
$\widetilde{c}_k$, energy parameter $\epsilon$, and coefficients $v_i$,
depend on the parameters $\omega$, $\lambda$ and $\mu$.  For simplicity, in
eqs~\eqref{27} this dependence has not been shown explicitly.  All
coefficients $\widetilde{c}_k$ depend on $\ell$ and on $v_{-1}$.
The coefficients $\widetilde{c}_k$ with $k\ge 3$ depend on $v_1$.  In
general, $v_i$ shows up in $\widetilde{c}_k$ with $k\ge (i+2)$ \cite{jk+as}. 
The external potential is proportional to $r^2$ and vanishes at $r=0$. 
Therefore, in expansion~\eqref{26} the $\omega$ dependence begins at
$\widetilde{c}_4$.

For the construction of $\widetilde{c_k}$ with even values of $k$, the state
energy is needed.  In these cases we use the expectation value of the
Hamiltonian defined in eq~\eqref{05}:
\[
E(\omega,\lambda,\mu)\,\approx\,E(\omega,\mu)+
\lambda\,\langle\bar{w}(r;\omega,0,\mu)\rangle,
\]
where $E(\omega,\mu)=\left.E(\omega,\lambda,\mu)\right|_{\lambda=0}$.
For definition of $\langle\bar{w}(r;\omega,\lambda,\mu)\rangle$ see
eq~\eqref{28}.  Numerical tests using the exact $E(\omega,\lambda,\mu)$
have shown that this approximation is negligible in comparison to the other
approximations made in the present paper.

\subsection{Dependence on $\lambda$ and normalization}
\label{ss-norm}

The GCC provide only ratios
$c_k(\mathfrak{p})/c_0(\mathfrak{p})\equiv\widetilde{c}_k(\mathfrak{p})$.  If
$\psi_\ell(r;\mathfrak{p})$ is known in the whole range of $r$ then
$c_0(\mathfrak{p})$ can be determined by the normalization condition.  In our
case this approach is nonapplicable since only the small-$r$ part of
$\psi_\ell(r;\mathfrak{p})$ is defined by the cusp conditions.  But
$c_0(\mathfrak{p})$ appears as a prefactor in the approximation~\eqref{26} for
$\psi_\ell(r;\mathfrak{p})$, and its value is necessary for any practical use
of this approximation.  Therefore, $c_0(\mathfrak{p})$ has to be estimated in
a different way using only the information about the short-range behavior of
the wave function.

We have to introduce additional information to deal with this issue.  Let us
first consider the dependence of $c_0$ on $\lambda$.  It is needed for the
adiabatic connection expression [eqs~\eqref{11}~and~\eqref{15}].  We
select $\left\|\psi_\mathrm{cm}(\mathbf{r}^+;\omega)\right\|=1$.  Then the
substitution of the wave function defined in eqs~\eqref{21} and \eqref{22}, 
and of its expansion~\eqref{26}, to eq~\eqref{14} yields 
\begin{eqnarray}
\nonumber
\langle\bar{w}(\mathfrak{p})\rangle&=&
\int_0^{2\pi}\int_0^\pi\int_0^\infty\,
\left|\psi_\mathrm{rel}(\mathbf{r};\mathfrak{p})\right|^2\,
\bar{w}(r;\mu)\,r^2dr\sin\theta\,d\theta\,d\phi\\
\label{28}
&=&\int_0^\infty\,\left|\psi_\ell(r;\mathfrak{p})\right|^2\,
\bar{w}(r;\mu)\,r^2dr,\\
&\approx&\left|c_0(\mathfrak{p})\right|^2\int_0^\infty\left|
\nonumber
r^{\ell+1}\sum_{k=0}^K\,\widetilde{c}_k(\mathfrak{p})
r^k\right|^2\,\bar{w}(r;\mu)\,dr,
\end{eqnarray}
and, according to eq~\eqref{11}, 
\begin{equation}
\label{29}
\thickbar{E}(\omega,\mu)=\int_0^1 
\langle\bar{w}(\omega,\lambda,\mu)\rangle\,d\lambda.
\end{equation}
For models close to the exact interaction (for large $\mu$) one can derive
the following relationship~\cite{GorSav-PRA-06, Sav-JCP-20} (a derivation
is given in the Appendix)
\begin{equation}
\label{30}
c_0(\omega,\lambda,\mu)\mathrel{\raisebox{-6pt}{$
\stackrel{\displaystyle{\sim}}
{\scriptstyle{\mu \to \infty}}$}}
\mathcal{N}\left(1+\frac{1-\lambda}
{\sqrt{\pi}\mu}+O\left(\mu^{-2}\right)\right), 
\end{equation}
where $\mathcal{N}$ is a still unknown normalization constant.
As one can see,
\[
\mathcal{N}=c_0(\omega,1,\mu)=c_0(\omega,\lambda,\infty)\equiv
c_0(\omega).
\]
Notice that $c_0(\omega,1,\mu)$ and $c_0(\omega,\lambda,\infty)$ do not
depend, respectively, on $\mu$ and on $\lambda$ and are 
equal to $c_0(\omega)$ corresponding to the physical (Coulomb) 
interaction potential.

We introduce the notation
\begin{equation}
\label{31}
\mathcal{I}_K(\omega,\lambda,\mu)=
\left(1+\frac{1-\lambda}{\sqrt{\pi}\mu}\right)^2 
\int_0^\infty\,\left[r^{\ell+1}\sum_{k=0}^K 
\widetilde{c}_k(\omega,\lambda,\mu)r^k\right]^2\,\bar{w}(r,\mu)dr.
\end{equation}
As stated at the beginning of this article, we know the model wave function 
at $\lambda=0$. Therefore, we can calculate
\begin{equation}
\label{32}
\langle\bar{w}(\omega,0,\mu)\rangle=\int_0^\infty
|\psi(r,\omega,0,\mu)|^2\bar{w}(r,\mu)r^2dr.
\end{equation}
Since $\mathcal{N}$ does not depend on $\lambda$, we can assume that
\begin{equation}
\label{33}
\mathcal{N}^2\approx\,\frac{\langle\bar{w}(\omega,0,\mu)\rangle}
{\mathcal{I}_K(\omega,0,\mu)}.
\end{equation}
Combining eqs~\eqref{29},~\eqref{28}, and~\eqref{33} we get
\begin{equation}
\label{34}
\thickbar{E}(\omega,\mu)\,\approx\,\frac{\langle\bar{w}(\omega,0,\mu)\rangle}
{\mathcal{I}_K(\omega,0,\mu)}\,
\int_0^1\mathcal{I}(\omega,\lambda,\mu)d\lambda
\end{equation}

\subsection{Corrections to the model}
\label{ss-corrmod}
Assume that $\omega$ is fixed and $\lambda=0$.  For simplicity, in
the next equations, we skip these parameters.  Also the $r$-dependence, in
most cases obvious, is not shown explicitly.  We designate $\mu$-dependence
only, retaining the convention that we drop $\mu$, if $\mu=\infty$. Our
model is defined by the analog of eq~\eqref{01}:
\begin{equation}
\label{35}
H(\mu)\psi(\mu)=E(\mu)\psi(\mu),\;\;\;\;\mbox{i.e.}\;\;\;\;
E(\mu)=\langle\psi(\mu)|H(\mu)|\psi(\mu)\rangle.
\end{equation}
For the physical (Coulombic) interaction, $\mu=\infty$ and we have 
$E=\langle\psi|H|\psi\rangle$. 

As we already said, the solutions of the crude model, $E(\mu)$ and
$\psi(\mu)$, are known.  The error of this model, $\Delta_\mathrm{0}E(\mu)$,
has been defined in eq~\eqref{03}.  Our aim is to add corrections to the
model.  The first-order perturbation correction to $E(\mu)$ is
\begin{equation}
\label{36}
\bra{\psi(\mu)}H-H(\mu)\ket{\psi(\mu)}=
\langle\psi(\mu)|\bar{w}(\mu)|\psi(\mu)\rangle.
\end{equation}
From here we have the correction:
\begin{equation}
\label{37}
\Delta_\mathrm{H}E(\mu)=\bra{\psi(\mu)}H\ket{\psi(\mu)}-E=E(\mu)-E+
\langle\bar{w}(\mu)\rangle.
\end{equation}

\begin{figure}[htb]
 \begin{center}
   \includegraphics[width=0.95\textwidth]{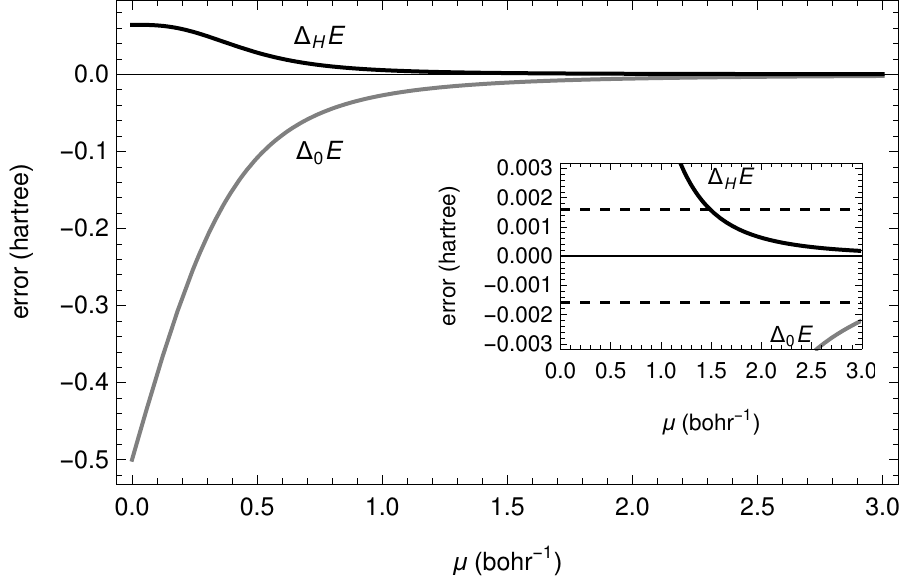}
 \end{center}
 \caption{Errors of the model energy, $\Delta_\mathrm{0}E(\mu)$, eq~\eqref{03},
 gray curve, and  of the expectation value of the physical Hamiltonian, $H$,
 with the model wave function, $\psi(\mu)$, $\Delta_\mathrm{H}E(\mu)$,
 eq~\eqref{37}, black curve, for the ground state of harmonium with
 $\omega=0.5$.  The inset zooms on the same curves (by two orders of
 magnitude), the horizontal dashed lines indicating the region of "chemical
 accuracy" ($\pm 1$~kcal/mol).}
 \label{fig:rangeoferrors}
\end{figure}

Plots of $\Delta_\mathrm{0}E(\mu)$ and $\Delta_\mathrm{H}E(\mu)$ are given
in Figure~\ref{fig:rangeoferrors}.  At $\mu=0$, that is for the
noninteracting system, $\Delta_\mathrm{0}E(\mu)$ is huge.  For the ground
state of harmonium with $\omega=1/2$, $E(\mu=0)=0.75$ instead of
$E=1.25$~hartree, giving $\Delta_\mathrm{0}E(\mu)=0.5$ hartree.  But,
$\Delta_\mathrm{0}E(\mu)$ does not fall to the chemical accuracy error bars
in the entire range of $\mu$ -- up to 3 $\mathrm{bohr}^{-1}$.  The
improvement due to the first-order perturbation (without any reference to a
specific structure of the wave function) is impressive.  For
$\Delta_\mathrm{H}E(\mu)$ the chemical accuracy is reached if
$\mu\,>\,1.5\,\mathrm{bohr}^{-1}$.  The error at $\mu=0$ is reduced from
0.5~hartree to $\sim 0.06$~hartree, that is of the same order of magnitude
as obtained with mean-field approximations ($\sim 0.04$~hartree for
Hartree-Fock or Kohn-Sham).  The sign of the errors in
Figure~\ref{fig:rangeoferrors} is (i) negative for $E(\mu)-E$ (as $w(r,\mu)
\le 1/r$), and (ii) positive for $\bra{\psi(\mu)}H\ket{\psi(\mu)}-E$ 
(by the variational principle).

Eq~\eqref{34} may be rewritten as
\begin{equation}
\label{38}
\Delta_\mathrm{K}E(\mu)=E(\mu)-E+C_\mathcal{I}^K\,
\langle\bar{w}(\mu)\rangle,
\end{equation}
where
\begin{equation}
\label{39}
C_\mathcal{I}^K\,=\,\frac{1}{\mathcal{I}_K(\omega,0,\mu)}\,
\int_0^1\mathcal{I}(\omega,\lambda,\mu)d\lambda.
\end{equation}
Equation~\eqref{38} can be interpreted as a generalization of eqs~\eqref{03}
and~\eqref{37}.  The prefactor of $\langle\bar{w}(\mu)\rangle$ in these two
equations is, respectively, $0$ and $1$.  Prefactor $C_\mathcal{I}^K$ in
eq~\eqref{38} has been derived using some information about the structure
of the wave function, specifically, GCC and the adiabatic connection. 
Therefore, one can expect that
\begin{equation}
\label{40}
\left|\Delta_\mathrm{0}E(\mu)\right|\,\gg\,
\left|\Delta_\mathrm{H}E(\mu)\right|\,
\gg\,\left|\Delta_\mathrm{K}E(\mu)\right|.
\end{equation}
The correctness of this expectation is demonstrated in the next section.

The integrals over $r$ in eq~\eqref{31} can be computed 
analytically~\cite{ryzhik}
\begin{equation}
\label{41}
\int_0^\infty\,\erfc(\mu\,r)r^k\,dr=
\frac{k\,\Gamma(k/2)}{2\sqrt{\pi}(k+1)\mu^{k+1}}.
\end{equation}
Therefore $\mathcal{I}_K(\omega,\lambda,\mu)$ can be expressed as an
algebric function of $\omega$, $\lambda$, and $\mu$.  Explicit formulas for
$\widetilde{c}_k(\mathfrak{p})$ can be easily deduced from eq~\eqref{27}. 
Consequently, the effort for computing the prefactor $C_\mathcal{I}^K$ in
eq~\eqref{38} is negligible.  Notice that in eq~\eqref{41}) $\Gamma(k/2)$
introduces a fast increase ot the absolute value of
$\mathcal{I}_K(\omega,\lambda,\mu)$ with $K$ (the maximum power of $r$ in
the expansion).

The approximation is valid only asymptotically, for sufficiently large
$\mu$.  Notice that expressions for $c_0(\mathfrak{p})$ and
$\mathcal{I}(\mathfrak{p})$ given in eqs~\eqref{30}, and~\eqref{31}, are
divergent at $\mu=0$.

\section{Results}
\label{s-results}
\subsection{General considerations}
\label{ss-gencons}
In the following we will consider only harmonium systems.  Of course, this
can be seen as futile, because solving the one-dimensional differential
Schr\"odinger equation, eq~\eqref{23}, is trivial (to obtain accurate
results we did it on a grid of the order of $10^5$ points). The choice of
this simple systems is motivated by the desire to study uniquely the effect
of the approximations introduced without adding other effects such as the
dependence on the one-particle basis set, or the expansion in terms of
Slater determinants.  To avoid introducing new effects, the one-particle
(external) potential is independent of $\mu$.  In view of future practical
applications, we are interested to see how weak the model interaction can be
taken (how small $\mu$ can be chosen), and still obtain an estimate of the
energy within "chemical accuracy".

We now define the "smallest acceptable $\mu$", SA$\mu$: for $\mu \ge
$~SA$\mu$ the absolute errors are smaller than the chemical accuracy
(1~kcal/mol).  The range of errors between $\pm1$~kcal/mol is indicated in
the plots by horizontal dashed lines.  For the system considered in
Figure~\ref{fig:rangeoferrors}, $E(\mu)$ has a SA$\mu$ slightly above
3~bohr$^{-1}$.  So, it is not interesting to discuss corrections if the
interaction already reaches this strength.  Weaker interactions can be
considered if we simply correct the energy to first order -- with
$\bra{\psi(\mu)}H\ket{\psi(\mu)}$, SA$\mu \approx 1.5$.

In the following, the presented plots show the errors of the approximations
of $E$ as functions of $\mu$ using the same scale as the inset in
Figure~\ref{fig:rangeoferrors}.

\begin{figure}
 \begin{center}
   \includegraphics[width=0.84\textwidth]{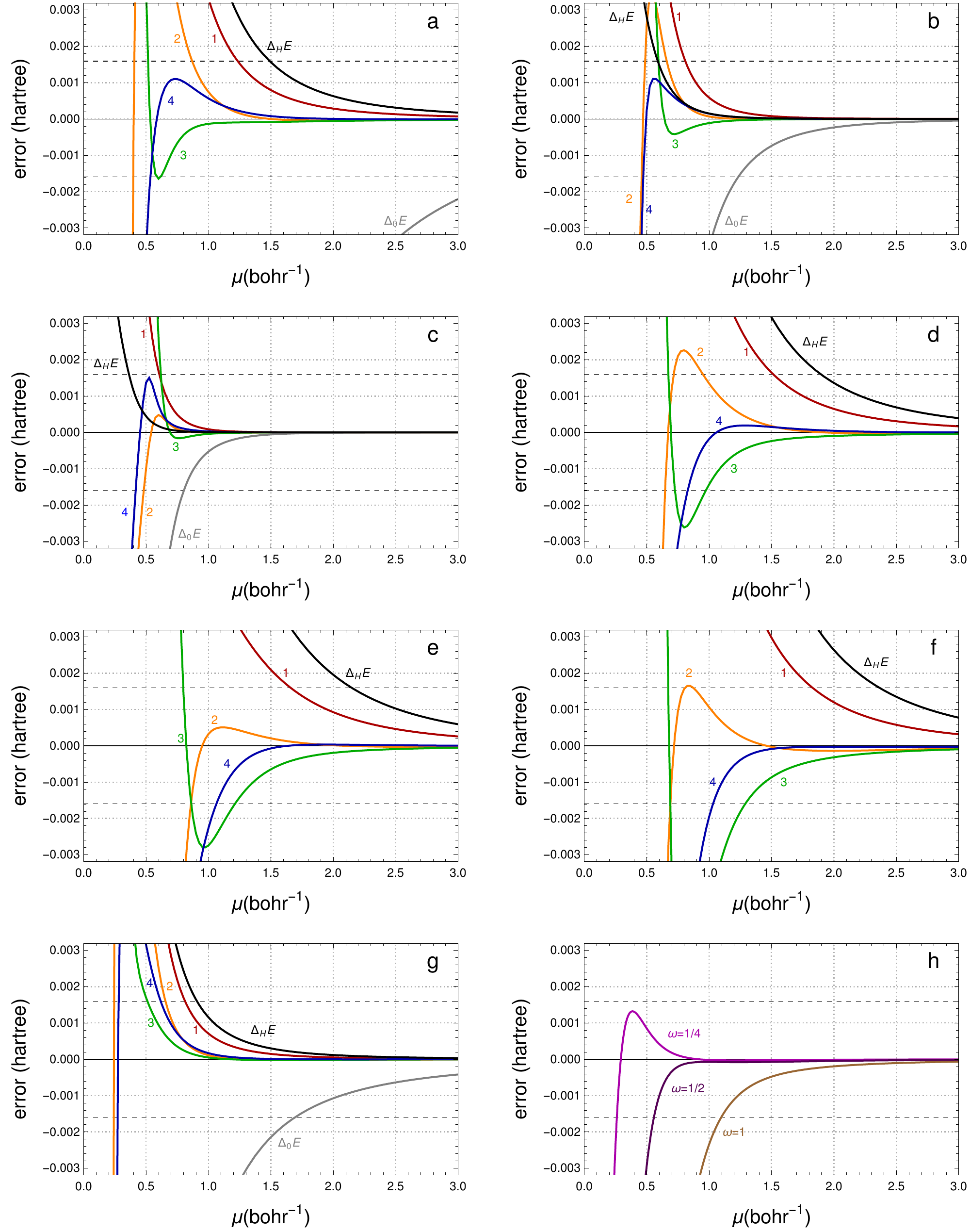}
 \end{center}
 \caption{Errors of the different approximations for the energy of harmonium
 systems.  The errors of the model energy, $\Delta_\mathrm{0}E(\mu)$, eq~\eqref{03}
 are represented as gray curves, that of the expectation value of the
 physical Hamiltonian 
 with the model wave function,
 $\Delta_\mathrm{H} E(\mu)$, eq~\eqref{37}, by black curves.  Panels a -- g show the
 errors for different orders $K$ in the GCC expansion, eq~\eqref{26}, the
 value of $K$ being indicated next to the curve.  Panel h shows the error of
 the approximation of order $\mu^{-4}$, as given in eq~\eqref{42}.
 Chemical accuracy ($\pm 1$~kcal/mol) is indicated by horizontal dashed lines.
a: $\omega=1/2, n=1, \ell=0$; 
b: $\omega=1/2, n=1, \ell=1$;
c: $\omega=1/2, n=1, \ell=2$;
d: $\omega=1/2, n=2, \ell=0$;
e: $\omega=1/2, n=3, \ell=0$;
f: $\omega=1, n=1, \ell=0$;
g: $\omega=1/4, n=1, \ell=0$;
h: $\omega=1/4, 1/2, \, \mathrm{or} \, 1$, and $ n=1, \ell=0$.
}
\label{fig:collectedfigs}
\end{figure}

\begin{figure}
   \centering
   \includegraphics[width=.9\textwidth]{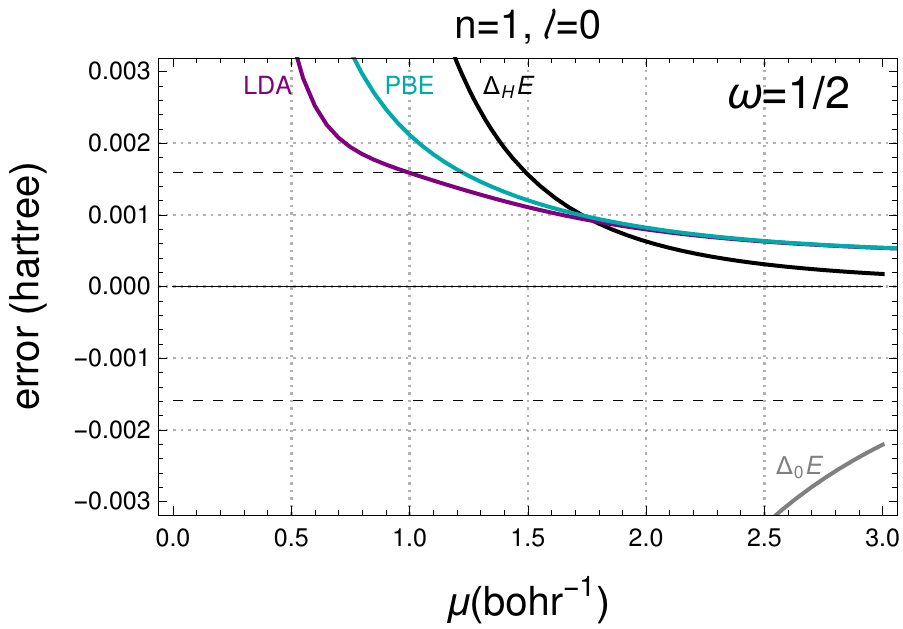} 
   \caption{Errors in the ground state energy estimate for harmonium
($\omega=1/2$) using density functional approximations ($\mu$-LDA and
$\mu$-PBE).  Also shown are the errors of the model energy, 
$\Delta_\mathrm{0}E(\mu)$, gray curve, and of the expectation value of 
the physical Hamiltonian, $\Delta_\mathrm{H} E(\mu)$, black curve.}
\label{fig:dfas}
\end{figure}

\subsection{How weak can the interaction be?}
\label{ss-howeak}

Let us consider the ground state of harmonium with $\omega=1/2$.  Now we use
the GCC to correct the model energy, $E(\mu)$ as in eqs~\eqref{34}
and~\eqref{38}, and cut off the expansion of the wave function
$\psi(r,\mathfrak{p})$, limiting the expansion in eq~\eqref{26} to a
maximal power of $r$, $K=1,2,3,4$ -- see Figure~\ref{fig:collectedfigs}a. 
When $K=1$ we satisfy only Kato's cusp condition.  This improves over
$E(\mu)$, and also over first-order perturbation theory, bringing the
SA$\mu$ to $1.3$~bohr$^{-1}$.  Increasing $K$ further reduces the SA$\mu$,
until a "wall" at around $0.5$~bohr$^{-1}$ is reached.  Notice that not only
the error is reduced by increasing $K$, but also the stability with respect
to the change of the results by chaning $\mu$ is increased when going beyond
satisfying only the Kato cusp condition.  This is important because (i) in
practice, there is an arbitrariness in the choice of $\mu$, and (ii) the
SA$\mu$ is system-dependent, as will be illustrated further down.  The
latter is of importance for size-consistency.

The derivations presented above never supposed that the state considered
corresponds to the ground state.  So, let us now consider some excited
states.  We consider first the lowest energy states with $\ell=1$ and
$\ell=2$.  The first corresponds to a triplet state,
(Figure~\ref{fig:collectedfigs}b), the second to a non-natural singlet
state~\cite{KutMor-JCP-92} (Figure~\ref{fig:collectedfigs}c); for $\ell=2$ the
singlet cusp condition does not give a prefactor $(1+r/2$) but $(1+r/6)$. 
King~\cite{Kin-96} remarks that, in an orbital picture, the $\ell=2$ state
corresponds to a strong mixture of sd and p$^2$ configurations.  Notice that
both the values provided by the model, and the expectation value of the
physical Hamiltonian, are now in much better agreement with the exact value
than for $\ell=0$: the prefactor $r^\ell$ appearing in eq~\eqref{26}
already keeps the electrons apart.  Even for $K=4$, for this system and
these states, there is no significant gain over using just 
$\bra{\psi(\mu)}H\ket{\psi(\mu)}$.  For $\ell=2$, the SA$\mu$s obtained 
for $K \le 4$ are not improved over that of the expectation value of 
the Hamiltonian.

We also present the first two excited states with $\ell=0$,
Figure~\ref{fig:collectedfigs}d,e.  The model system has much larger errors
which fall outside the domain of the plots.  We notice an overall worsening
ot the quality of the approximations.  The model shows errors that fall
outside the range of the plots.  The SA$\mu$ for the expectation value of
the Hamiltonian is around 2~bohr$^{-1}$.  Using the Kato cusp condition
reduces it to $\approx 1.5$~bohr$^{-1}$.  Increasing the order of the
expansion, $K$, moves the SA$\mu$ down to somewhere between 1 and
0.5~bohr$^{-1}$.

We have seen above that the SA$\mu$s can be quite sensitive to the state
described.  They can also be sensitive to the system.  Modifying $\omega$
affects the SA$\mu$s.  Figure~\ref{fig:collectedfigs}f,g show the effect
of making the system more compact ($\omega=1$~au), or diffuse
($\omega=1/4$~au), for the respective ground states.

\subsection{Comparisons}
\label{ss-compar}
Using the GCC is not the only way to describe the short-range behavior.  One
can replace the expansion in powers of $r$, eq~\eqref{26} by a simple form
that does not diverge as $r \rightarrow \infty$.  For example, one can
ignore all terms arising from the external potential and the energy in the
Schr\"odinger equation, in the limit $\mu \rightarrow \infty$, and we get,
to order $\mu^{-4}$ (see ref~\citenum{GorSav-PRA-06}, and eq~(21)
in ref~\citenum{Sav-JCP-20}),
\begin{equation}
\label{42}
    \frac{\int_0^1 d \lambda \, 
    \mathcal{I}(\lambda,\mu)}{\mathcal{I}(\lambda=0,\mu)} =
    \frac{\mu^2+1.06385 \mu + 0.31982}
         {\mu^2 + 1.37544\mu + 0.487806 }
\end{equation}
A comparison of Figures~\ref{fig:collectedfigs}a,~\ref{fig:collectedfigs}f,
or~\ref{fig:collectedfigs}g to~\ref{fig:collectedfigs}h shows that this
approximation performs very well compared to those obtained from the GCC:
the "walls" are at comparable values of $\mu$.  This is encouraging because
the GCC require, in general, taking into account the external potential and
the energy.

Density functional approximations have been used for many years to correct
models for missing short-range interaction.~\cite{SavFla-IJQC-95,Sav-INC-96}. 
Figure~\ref{fig:dfas} shows the corrections provided for two approximations,
the local density approximation, LDA, and that of Perdew, Burke, and
Ernzerhof, PBE, modified to depend on $\mu$.~\cite{Sav-INC-96,
PazMorGorBac-PRB-06, GolWerSto-PCCP-05} The SA$\mu$s are around
1~bohr$^{-1}$, thus improving over using only Kato's cusp condition (cf. 
Figure~\ref{fig:collectedfigs}a).  It may be at first surprising that in
Figure~\ref{fig:dfas} LDA is slightly better than PBE.  However, PBE becomes
better than LDA for small $\mu$.  This cannot be seen in Figure~\ref{fig:dfas}
because it shows only the region of "chemical accuracy"; the errors for
small $\mu$ are much larger.

\section{Conclusions}
\label{s-concl}

\subsection{Summary}
\label{ss-sum}

We have considered model systems, where electrons interact only via a
long-range potential, eq~\eqref{02}.  In order to obtain the physical
energy we explored corrections based upon the short-range behavior of the
wave function, both for the ground and excited states.  The numerical
results were all obtained for harmonium, eqs~\eqref{16}~--~\eqref{22} where
accurate results are easy to reach due to separability.
 
We are interested in having corrections to models where the interaction is
very weak.  However, the approximations proposed can be systematically
improved in an asymptotic sense: as the order of the approximation
increases, there is a domain of models close enough to the exact solution
that gets systematically better.  Unfortunately, at present, the
approximations fail when the range of the interaction between particles
becomes too large (our parameter $\mu$ becomes too small).  We attribute it
to imposing only the short-range behavior of the wave function.  For short
range, the results turned out to be more reliable than obtained by
correcting the models with density functional approximations. 
Unfortunately, the range of validity is system- and state-dependent.

\subsection{Perspectives}
\label{ss-persp}

In quantum chemistry there are different approaches to tackle the problem
raised by the singularity of the Coulomb potential.  A ``brute force''
pathway, is to use a large expansion in Slater determinants.  The burden can
be reduced significantly by using ``selected configuration interaction''
techniques (see, e.g., ref \citenum{ChilkNes-21} and references therein). 
Another way is to improve the description of the wave function by the use of
correlation factors, like Jastrow factors in Quantum Monte Carlo ({\tt see,
e.g., refs \citenum{JastR-55, LuStSchMoo-15, PetTouUm-12, BoThAla-09}}), or F12
methods (see, e.g., refs \citenum{Ten-TCA-12,TewKlo-MP-10}). Still another
way to approach this problem is to use density functionals that transfer the
short-range behavior from other systems (as a rule, from the uniform
electron gas).

In this paper, we combine the spirit of the last two approaches.  As in
density functional calculations, we compute a model system for which the
interaction has no singularity (and thus is expected to converge faster, in
general).  However, in contrast to density functional calculations, no
density functional approximations are present, and the Hohenberg-Kohn
theorem (formulated for ground states) is not used.  The approach is
applicable to ground and excited states.

As in methods using correlation factors we use the exact short-range
behavior of the wave function.  The trick allowing to use it comes from
having a correction that depends only on the missing part of the interaction
that is short-ranged.

The present paper did only show exploratory calculations.  However, it is
possible to extend considerations presented in this paper to other
systems.  Already Kurokawa {\em et al.}~\cite{kurokawa2} have shown that the
GCC can be applied when the Schr\"odinger equation for the relative
coordinate cannot be separated from that of the center of mass (the He
atom).

Before extending our approach there are several issues to be explored. 
Probably the first one is its formulation in terms of reduced density
matrices.  In this paper we use GCC for the wave function.  To generalize
our approach, it might be useful to express the GCC in terms of the
2RDM~\cite{nagy1,nagy2}.  

One may ask whether it is not more convenient to re-express our formulation
in terms of the 1RDM.  In 1975 Kimball found a relationship between the 2RDM
at coalescence and the distribution of momentum, in fact, the behavior of
the 1RDM \cite{Kimb-75}.  At the same time Yasuhara demonstrated that the
energy of the uniform electron gas can be expressed in terms of the kinetic
energy (one-particle operators) instead of the two-body interations
\cite{Yasu-75}.  This gave rise to studies on using the adiabatic connection
on the kinetic energy (i.e.  the 1RDM) rather than on two-body interactions
\cite{Sav-95, LeGor-95, TeaHelSav-16}.  Recently, the effects of the
electron-electron coalescence on the properties of the natural orbitals and
structure of the 1RDM have been revealed \cite{Cio-21-a, Cio-21-b}.

Another issue is the dependence of the GCC on the coalescence
point~\cite{kurokawa1,kurokawa2,kurokawa3,jk+as}.  For this, we can get
inspiration from density functional approximation: in each point of space
one has a different approximation. Even the Kato's cusp condition (that
seemingly is universal) contains a state dependence (through $\ell$).  In
density functional calculation it is treated by using the spin polarization
\cite{BeSaSt-95}, but it can be treated in the context of a pair density
(see, e.g., ref \citenum{PerSaBu-95}).

We illustrate the problematic with two two-electron harmonium systems
($\mathrm{A}$ and $\mathrm{B}$) that instead of being treated separately are
treated together (for example, having two quantum dots). The system
may be described by a double-well potential with two wells
sufficiently separated and deep, so that each of them can be approximated by
a harmonic potential.  We assume that this separation is large enough to
neglect the exchange effects between electrons in $\mathrm{A}$ and
$\mathrm{B}$.  Consequently, the spin part may be separated and, in effect, 
we can cosider the total wave function which depends on the orbital variables 
only. It can be written as a product of two harmonium wave functions: 
\begin{equation}
\label{43}
\Psi_\mathrm{total}(\mathbf{r}_1,\mathbf{r}_2,\mathbf{r}_3,\mathbf{r}_4)= 
\Psi_\mathrm{A}(\mathbf{r}_1,\mathbf{r}_2)
\Psi_\mathrm{B}(\mathbf{r}_3,\mathbf{r}_4)
\end{equation}
Let us first consider that subsystem $\mathrm{A}$ is in a state with
$\ell=0$, while subsystem $\mathrm{B}$ is in a state with $\ell=1$.  In
region $\mathrm{A}$ the coefficient $\tilde{c}_1 = 1/2$, while in region
$\mathrm{B}$, $\tilde{c}_1 = 1/4$.  Let us now consider another example.  In
both subsystems we have $\ell=0$.  However, the values of $\omega$ differ:
$\omega_\mathrm{A} \ne \omega_\mathrm{B}$.  We can see from
Figures~\ref{fig:collectedfigs}a,~\ref{fig:collectedfigs}f-h that the range of
where results are reliable depend on the model chosen.  For example, we see
in Figure~\ref{fig:collectedfigs}h that if we choose $\mu=0.5$ we will have
chemical accuracy for a subsystem with $\omega=1/4$, but that the error is
significantly larger for a subsystem with $\omega=1$ where chemical accuracy
is reached only for $\mu=1$.  Do we choose to make the more expensive global
calculation, say with $\mu=1$, or do we decide to lose quality and make the
cheaper calculation at $\mu=1/2$?  Alternatively, we might consider making
calculations with models that change from one region of space to another.

In how far do we need to go beyond the Kato cusp condition?  Using
eq~\eqref{42}, Figure~\ref{fig:collectedfigs}h, shows that this might not be
needed.  Furthermore, eq~\eqref{a7} of the Appendix (see also discussion in
ref~\citenum{GorSav-PRA-06}) shows that there are terms that are important
when $r > 1/\mu$ even when $r$ is small and they may need a more careful
treatment.

Another question is raised by analyzing the dependence on the external
potential.  It was shown by Kurokawa {\it et al.}~\cite{kurokawa2} that for
small interelectronic distance $r$, the leading term in the expansion of
the Coulomb potential of the nuclei is quadratic.  So, some of the
conclusions drawn from the present treatment of harmonium might be 
applied in systems where the external potential is Coulombic.

\section{Appendix: A derivation of eq~\eqref{30}}
\label{s-appendix}

We repeat here the argumentation given in ref~\citenum{GorSav-PRA-06},
section III, where the result was obtained for $\lambda=0$.

We consider the behavior of the Schr\"odinger equation
\begin{equation}
\left (-\frac{d^2}{dr^2} - \frac{2}{r}\,\frac{d}{d r} + 
\frac{(1-\lambda)\,\mathrm{erf}(\mu\,r)+\lambda}{r}+\frac{\omega^2r^2}{2}
-E\right)\psi(r)=0
\end{equation}
at the limit of large $\mu$. After a change of the variable $r$ to 
$x=\mu r$, and defining $u(x) = \psi(x/\mu)$ one obtains 
\begin{equation}
\left (-\frac{d^2}{dx^2} - \frac{2}{x}\,\frac{d}{d x} + 
\frac{(1-\lambda)\,\mathrm{erf}(x)+\lambda}{\mu\,x}+
\frac{\omega^2x^2}{2\,\mu^4}-\frac{E}{\mu^2}\right)u(x)=0.
\end{equation}
For large $\mu$, we neglect terms proportional to $(1/\mu)^n$ with $n>1$
and to the resulting equation apply perturbation theory up to the first
order in $1/\mu$. We set:
\begin{equation}
u(x) = u^{(0)}(x) + \frac{1}{\mu} u^{(1)}(x).
\end{equation}
The $0$th order in $1/\mu$ gives
\begin{equation}
\left (-\frac{d^2}{dx^2} - \frac{2}{x}\,\frac{d}{d x}\right)u^{(0)}(x)=0.
\end{equation}
From here we get 
\begin{equation}
u^0(x) = \mathcal{A}_0 + \frac{\mathcal{A}_1}{x}
\end{equation}
where $\mathcal{A}_0$ and $\mathcal{A}_1$ are integration constants. 
The wave function has to be finite at $x=0$.  Therefore,
$\mathcal{A}_1=0$.  Furthermore, as $u^{(0)}$ corresponds to the
solution at $\mu=\infty$, namely to the exact solution independent of
either $\lambda$ or $\mu$, we set $\mathcal{A}_0=\mathcal{N}$.
Consequently, the equation for $u^{(1)}(x)$ is
\begin{equation}
\left (-\frac{d^2}{dx^2}-\frac{2}{x}\,\frac{d}{d x}\right)u^{(1)}(x)+ 
\frac{(1-\lambda)\,\mathrm{erf}(x)+\lambda}{x}\,\mathcal{N}=0.
\end{equation}
After solving this equation and changing variable $x$
to $r$, we get
\begin{equation}
\label{a7}
\psi(r) = \mathcal{N} \left ( 1 + \frac{\lambda}{2} r +
\frac{1-\lambda}{2 \sqrt{\pi} \mu } e^{- \mu^2 r^2} + (1-\lambda)
\left(\frac{r}{2} + \frac{1}{4 \, \mu^2 \, r} \right) \mathrm{erf}(\mu r) 
\right)+\mathcal{B}_0 + \frac{\mathcal{B}_1}{\mu r},
\end{equation}
where $\mathcal{B}_0$ and $\mathcal{B}_1$ are integration constants.
To avoid singularity of the wave function at $r=0$ $\mathcal{B}_1=0$.
In order to recover results for $\lambda=0$ \cite{GorSav-PRA-06} we 
have to set $\mathcal{B}_0=0$. Finally, in the limit $r\,\rightarrow\,0$, 
up to the first-order in $1/\mu$, we have
\begin{equation}
\left.\psi(r)\right|_{r=0}=\mathcal{N}
\left(1+\frac{1-\lambda}{\sqrt{\pi}\mu}\right),
\end{equation}
where $\mathcal{N}$ is a normalization constant indepndent of either $\lambda$
or $\mu$. 

\section{Acknowledgement}
\label{s-acknowledgement}

This work was done without specific financial support.  It was presented in
part at the MQM 2022 honoring Profs.  Gustavo E.  Scuseria (Rice University,
Houston, Texas, U.S.A.) and Martin Head-Gordon (Univerisity of California,
Berkeley, California, U.S.A.) held at Virginia Tech (Blacksburg, Virginia,
U.S.A.).

\end{document}